%% file: bicmb_cp.tex
\documentclass[journal,12pt,onecolumn]{IEEEtran}

\usepackage{setspace}
\ifCLASSOPTIONonecolumn \doublespace \fi
\usepackage{graphics}
\usepackage{rotating}
\usepackage{amsfonts}
\usepackage{amsmath}
\usepackage[tight,footnotesize]{subfigure}
\usepackage{multirow}
\allowdisplaybreaks[1]
\newtheorem{theorem}{Theorem}
\renewcommand{\IEEEQED}{\IEEEQEDopen}

\graphicspath{{Figures/}} \DeclareGraphicsExtensions{.eps,.ps}

\begin{document}

\title{Bit-Interleaved Coded Multiple Beamforming with Constellation Precoding}

\ifCLASSOPTIONconference
\author{\IEEEauthorblockN{Hong Ju Park and Ender Ayanoglu}\\
\IEEEauthorblockA{Center for Pervasive Communications and Computing\\
Department of Electrical Engineering and Computer Science\\
University of California, Irvine\\
Email: hjpark@uci.edu, ayanoglu@uci.edu}}
\else
\author{Hong~Ju~Park ~\IEEEmembership{Student~Member,~IEEE,}
and~Ender~Ayanoglu ~\IEEEmembership{Fellow,~IEEE}\\
Center for Pervasive Communications and Computing\\
Department of Electrical Engineering and Computer Science\\
University of California, Irvine\\
Email: hjpark@uci.edu, ayanoglu@uci.edu}
\fi

\maketitle

\ifCLASSOPTIONonecolumn
 \setlength\arraycolsep{4pt}
\else
 \setlength\arraycolsep{2pt}
\fi

\input{abstract}
\input{introduction}
\input{system_model}
\input{div_analysis}
\input{results}
\input{conclusion}

\bibliographystyle{IEEEtran}
\bibliography{bicmb_cp.bbl}

\end{document}

%% file: abstract.tex
\begin{abstract}
In this paper, we present the diversity order analysis of
bit-interleaved coded multiple beamforming (BICMB) combined with the
constellation precoding scheme. Multiple beamforming is realized by
singular value decomposition of the channel matrix which is assumed
to be perfectly known to the transmitter as well as the receiver.
Previously, BICMB is known to have a diversity order bound related
with the product of the code rate and the number of parallel
subchannels, losing the full diversity order in some cases. In this
paper, we show that BICMB combined with the constellation precoder
and maximum likelihood detection achieves the full diversity order.
We also provide simulation results that match the analysis.
\end{abstract}

%% file: introduction.tex
\section{Introduction} \label{sec:introduction}

When the perfect channel state information is available at the
transmitter to achieve spatial multiplexing and thereby increase the
data rate, or to enhance the performance of a multi-input
multi-output (MIMO) system, beamforming can be employed
\cite{jafarkhaniBook}. The beamforming vectors are designed in
\cite{SampathJCOM01}, \cite{palomarTSP03} for various design
criteria, and can be obtained by singular value decomposition (SVD),
leading to a channel-diagonalizing structure optimum in minimizing
the average bit error rate (BER) \cite{palomarTSP03}.

It is known that an SVD subchannel with larger singular value
provides larger diversity gain. During the simultaneous parallel
transmission of the symbols on the diagonalized subchannels, the
performance is dominated by the subchannel with the smallest
singular value, resulting in losing the full diversity order
\cite{sengulTC06AnalSingleMultpleBeam}, \cite{OrdonezTSP07}. To
overcome the degradation of the diversity order of multiple
beamforming, bit-interleaved coded multiple beamforming (BICMB) was
proposed \cite{akayTC06BICMB}, \cite{akayTC06BICMB_arxiv}. This
scheme interleaves the codewords through the multiple subchannels
with different diversity orders, resulting in better diversity
order. BICMB can achieve the full diversity order offered by the
channel as long as the code rate $R_c$ and the number of subchannels
used $S$ satisfy the condition $R_c S \leq 1$ \cite{ParkICC09}.

We showed in \cite{park-2009_arxiv} and \cite{ParkGlobecom09} that
constellation precoded multiple beamforming, which converts a symbol
into a precoded symbol and distributes the precoded symbol over the
subchannels, can compensate for the diversity loss caused by the
uncoded multiple beamforming. In this paper, by calculating pairwise
error probability (PEP), we present the diversity analysis of
Bit-Interleaved Coded Multiple Beamforming with Constellation
Precoding (BICMB-CP), which adds the constellation precoding stage
to BICMB. We show that adding the constellation precoder to the
BICMB system which does not satisfy the full diversity condition
guarantees the full diversity order when the subchannels to transmit
the precoded symbols are properly chosen. Simulation results are
shown to prove the analysis.

The rest of this paper is organized as follows. The description of
BICMB-CP is given in Section \ref{sec:system_model}. Section
\ref{sec:div_analysis} presents the diversity analysis through the
calculation of the upper bound to PEP. Simulation results supporting
the analysis are shown in Section \ref{sec:results}. Finally, we end
the paper with our conclusion in Section \ref{sec:conclusion}.

\textbf{Notation:} Bold lower (upper) case letters denote vectors
(matrices). $\textrm{diag}[\mathbf{B}_1, \cdots, \mathbf{B}_P]$
stands for a block diagonal matrix with matrices $\mathbf{B}_1,
\cdots, \mathbf{B}_P$, and $\textrm{diag}[b_1, \cdots, b_P]$ is a
diagonal matrix with diagonal entries $b_1, \cdots, b_P$. The
superscripts $(\cdot)^H$, $(\cdot)^T$, $(\cdot)^*$, $\bar{(\cdot)}$
stand for conjugate transpose, transpose, complex conjugate, binary
complement, respectively, and $\forall$ denotes for-all.
$\mathbb{R}^+$ and $\mathbb{C}$ stand for the set of positive real
numbers and the complex numbers, repectively. $d_{min}$ is the
minimum Euclidean distance between two points in the constellation.
$N$ and $M$ stand for the number of transmit and receive antennas.

%% file: system_model.tex
\section{BICMB with Constellation Precoding} \label{sec:system_model}

Fig. \ref{fig:system_model} represents the structure of BICMB with
constellation precoding. First, the code rate $R_c = k_c / n_c$
convolutional encoder, possibly combined with a perforation matrix
for a high rate punctured code, generates the codeword $\mathbf{c}$
from the information bits. Then, the spatial interleaver distributes
the coded bits into $S \leq \min(N, M)$ streams, each of which is
interleaved by an independent bit-wise interleaver $\pi$. The
interleaved bits are mapped by Gray encoding onto the symbol
sequence $\mathbf{X} = [\mathbf{x}_1 \, \cdots \, \mathbf{x}_K]$,
where $\mathbf{x}_k$ is an $S \times 1$ symbol vector at the
$k^{th}$ time instant. In this model, we assume that each stream
employs the same modulation scheme. Each entry in the symbol vector
belongs to a signal set $\chi \subset \mathbb{C}$ of size $|\chi| =
2^m$, such as $2^m$-QAM, where $m$ is the number of input bits to
the Gray encoder.

\ifCLASSOPTIONonecolumn
\begin{figure}[!m]
\centering \includegraphics[width = 0.6\linewidth]{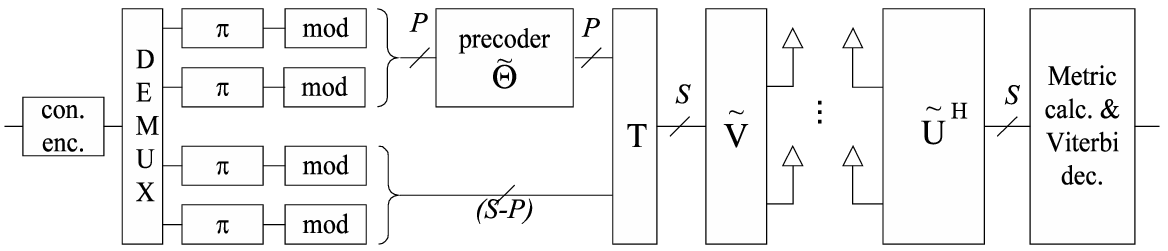}
\caption{Structure of Bit-Interleaved Coded Multiple Beamforming
with Constellation Precoding.} \label{fig:system_model}
\end{figure}
\else
\begin{figure}[!t]
\centering \includegraphics[width = 1.0\linewidth]{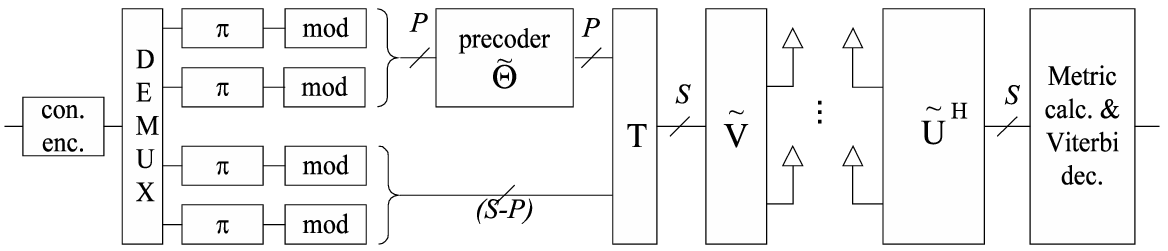}
\caption{Structure of Bit-Interleaved Coded Multiple Beamforming
with Constellation Precoding.} \label{fig:system_model}
\end{figure}
\fi

The symbol vector $\mathbf{x}_k$ is multiplied by the $S \times S$
precoder $\boldsymbol{\Theta}$, which is defined as
\begin{align}
\mathbf{\Theta} = \left[ \begin{array}{cc}
\mathbf{\tilde{\Theta}} & \mathbf{0} \\
\mathbf{0} & \mathbf{I}_{S-P}
\end{array} \right]
\label{eq:precoder_def}
\end{align}
where $\mathbf{\tilde{\Theta}}$ is the $P \times P$ unitary
constellation precoding matrix that precodes the first $P$ modulated
entries of the vector $\mathbf{x}_k$. When all of the $S$ modulated
entries are precoded ($P = S$), we call the resulting system
Bit-Interleaved Coded Multiple Beamforming with Full Precoding
(BICMB-FP), otherwise, we call it Bit-Interleaved Coded Multiple
Beamforming with Partial Precoding (BICMB-PP). The symbol generated
by $\boldsymbol{\Theta}$ is multiplied by $\mathbf{T}$ which is an
\mbox{$S \times S$} permutation matrix to define the mapping of the
precoded and non-precoded symbols onto the predefined subchannels.
Let us define $\mathbf{b}_p = \left[ b_p(1) \, \cdots \, b_p(P)
\right]$ as a vector whose element $b_p(u)$ is the subchannel on
which the precoded symbols are transmitted, and ordered increasingly
such that $b_p(u) < b_p(v)$ for $u < v$. In the same way,
$\mathbf{b}_n = \left[ b_n(1) \, \cdots \, b_n(S-P) \right]$ is
defined as an increasingly ordered vector whose element $b_n(u)$ is
the subchannel which carries the non-precoded symbols.

The MIMO channel $\mathbf{H} \in \mathbb{C}^{M \times N}$ is assumed
to be quasi-static, Rayleigh, and flat fading, and perfectly known
to both the transmitter and the receiver. In this channel model, we
consider that the channel coefficients remain constant for the $K$
symbol duration. The beamforming vectors are determined by the SVD
of the MIMO channel, i.e., $\mathbf{H} = \mathbf{U \Lambda V}^H$
where $\mathbf{U}$ and $\mathbf{V}$ are unitary matrices, and
$\mathbf{\Lambda}$ is a diagonal matrix whose $s^{th}$ diagonal
element, $\lambda_s \in \mathbb{R}^+$, is a singular value of
$\mathbf{H}$ in decreasing order. When $S$ symbols are transmitted
at the same time, then the first $S$ vectors of $\mathbf{U}$ and
$\mathbf{V}$ are chosen to be used as beamforming matrices at the
receiver and the transmitter, respectively. $\mathbf{\tilde{U}}$ and
$\mathbf{\tilde{V}}$ in Fig. \ref{fig:system_model} denote the first
$S$ column vectors of $\mathbf{U}$ and $\mathbf{V}$.

The spatial interleaver arranges the symbol vector $\mathbf{x}_k$ as
$\mathbf{x}_k = [\mathbf{x}_{k, \mathbf{b}_p}^T \, \vdots \,
\mathbf{x}_{k, \mathbf{b}_n}^T]^T = [x_{k,b_p(1)} \, \cdots \,
x_{k,b_p(P)} \, \vdots$ $\, x_{k,b_n(1)} \, \cdots \, x_{k,
b_n(S-P)}]^T$ where $\mathbf{x}_{k, \mathbf{b}_p}$ and
$\mathbf{x}_{k, \mathbf{b}_n}$ are the modulated entries to be
transmitted on the subchannels specified in $\mathbf{b}_p$ and
$\mathbf{b}_n$, respectively. The $S \times 1$ detected symbol
vector $\mathbf{r}_k = [ (\mathbf{r}_{k}^p)^T \, \vdots \,
(\mathbf{r}_{k}^n)^T]^T = [r_{k,1} \, \cdots \, r_{k, P} \, \vdots
\, r_{k, P+1} \, \cdots \, r_{k,S}]^T$ at the $k^{th}$ time instant
is
\begin{align}
\mathbf{r}_k = \boldsymbol{\Gamma} \mathbf{\Theta x}_k +
\mathbf{n}_k \label{eq:detected_symbol}
\end{align}
where $\boldsymbol{\Gamma}$ is a block diagonal matrix,
$\boldsymbol{\Gamma} = \textrm{diag}[\boldsymbol{\Gamma}_p ,
\boldsymbol{\Gamma}_n]$ with diagonal matrices defined as
$\boldsymbol{\Gamma}_p = \textrm{diag}[\lambda_{b_p(1)}, $ $\,
\cdots, \, \lambda_{b_p(P)}]$, $\boldsymbol{\Gamma}_n =
\textrm{diag}[\lambda_{b_n(1)}, \, \cdots, \, \lambda_{b_n(S-P)}]$,
and $\mathbf{n}_k = [ (\mathbf{n}_{k}^p)^T \, \vdots \,
(\mathbf{n}_{k}^n)^T]^T = [r_{n,1} \, \cdots \, r_{n, P} \, \vdots
\, r_{n, P+1} \, \cdots \, r_{n,S}]^T$ is an additive white Gaussian
noise vector with zero mean and variance $N_0 = N / SNR$.
$\mathbf{H}$ is complex Gaussian with zero mean and unit variance,
and to make the received signal-to-noise ratio $SNR$, the total
transmitted power is scaled as $N$. The input-output relation in
(\ref{eq:detected_symbol}) is decomposed into two equations as
\begin{equation}
\begin{split}
\mathbf{r}_{k}^p = \boldsymbol{\Gamma}_p \boldsymbol{\tilde{\Theta
}} \mathbf{x}_{k, \mathbf{b}_p} + \mathbf{n}_k^p \\
\mathbf{r}_{k}^n = \boldsymbol{\Gamma}_n \mathbf{x}_{k,
\mathbf{b}_n} + \mathbf{n}_k^n. \label{eq:deteced_symbol_decomposed}
\end{split}
\end{equation}

The location of the coded bit $c_{k'}$ within the symbol sequence
$\mathbf{X}$ is known as $k' \rightarrow (k, l, i)$, where $k$, $l$,
and $i$ are the time instant in $\mathbf{X}$, the symbol position in
$\mathbf{x}_k$, and the bit position on the symbol $x_{k,l}$,
respectively. Let $\chi_{b}^{i}$ denote a subset of $\chi$ whose
labels have $b \in \{0, 1\}$ in the $i^{th}$ bit position. By using
the location information and the input-output relation in
(\ref{eq:detected_symbol}), the receiver calculates the maximum
likelihood (ML) bit metrics for the coded bit $c_{k'}$ as
\begin{align}
\gamma^{l,i}(\mathbf{r}_{k}, c_{k'}) = \min_{\mathbf{x} \in
\xi_{c_{k'}}^{l,i}} \| \mathbf{r}_{k} - \boldsymbol{\Gamma}
\boldsymbol{\Theta} \mathbf{x} \|^2 \label{eq:ML_bit_metrics}
\end{align}
where $\xi_{c_{k'}}^{l,i}$ is a subset of $\chi^S$, defined as
\begin{align*}
\xi_{b}^{l,i} = \{ \mathbf{x} = [x_1 \, \cdots \, x_S ]^T :
x_{s|s=l} \in \chi_{b}^{i}, \textrm{ and } x_{s|s \neq l} \in \chi
\}.
\end{align*}
In particular, the bit metrics, equivalent to
(\ref{eq:ML_bit_metrics}) for partial precoding, are represented as
\begin{align}
\gamma^{l,i}(\mathbf{r}_{k}, c_{k'}) =  \left\{
\begin{array}{ll}
\min\limits_{\mathbf{x} \in \psi_{c_{k'}}^{l,i}} \| \mathbf{r}_{k}^p
- \boldsymbol{\Gamma}_p \boldsymbol{\tilde{\Theta}} \mathbf{x} \|^2,
& \textrm{ if $1 \leq l \leq P$} \\
\min\limits_{x \in \chi_{c_{k'}}^{i}} |r_{k,l} - \lambda_{\hat{l}} x
|^2, & \textrm{ if $P+1 \leq l \leq S$}
\end{array} \right.
\label{eq:ML_bit_metrics_PPMB}
\end{align}
where $\psi_{b}^{l,i}$ is a set which is mapped from the set
$\xi_b^{l,i}$ by a surjective function $f(\mathbf{x})$, for
$\mathbf{x} = [x_1 \, \cdots \, x_S]^T$, defined as
\begin{equation*}
f\left( \mathbf{x} \right) = [x_1 \, \cdots \, x_P]^T,
\end{equation*}
and $\hat{l}$ is an entry in $\mathbf{b}_n$, corresponding to the
subchannel mapped by $\mathbf{T}$. Finally, the ML decoder makes
decisions according to the rule
\begin{align}
\mathbf{\hat{c}} = \arg\min_{\mathbf{\tilde{c}}} \sum_{k'}
\gamma^{l,i}(\mathbf{r}_{k}, \tilde{c}_{k'}).
\label{eq:Decision_Rule}
\end{align}

%% file: div_analysis.tex
\section{Diversity Analysis} \label{sec:div_analysis}

Since BER in BICMB is bounded by the union of the PEP corresponding
to each error event \cite{akayTC06BICMB}, the calculation of each
PEP is needed. In particular, the overall diversity order is
dominated by the pairwise errors which have the smallest exponent of
signal-to-noise ratio in PEP representation. In this section, we
calculate the upper bound to each PEP corresponding to the pairwise
errors.

\subsection{BICMB with Full Precoding} \label{sec:FPMB}

Based on the bit metrics in (\ref{eq:ML_bit_metrics}), the
instantaneous PEP between the transmitted codeword $\mathbf{c}$ and
the decoded codeword $\mathbf{\hat{c}}$ is calculated as
\begin{align}
\textrm{Pr} \left( \mathbf{c} \rightarrow \mathbf{\hat{c}} |
\mathbf{H} \right) = \textrm{Pr} \left( \sum_{k'} \min_{\mathbf{x}
\in \xi_{c_{k'}}^{l,i}} \| \mathbf{r}_k - \boldsymbol{\Gamma}
\boldsymbol{\Theta} \mathbf{x} \|^2 \geq \right. \ifCLASSOPTIONtwocolumn \nonumber \\
\fi \left. \sum_{k'} \min_{\mathbf{x} \in \xi_{\hat{c}_{k'}}^{l,i}}
\| \mathbf{r}_k - \boldsymbol{\Gamma} \boldsymbol{\Theta} \mathbf{x}
\|^2 \right) \label{eq:PEP_original}
\end{align}
where $c_{k'}$ and $\hat{c}_{k'}$ is the coded bit of $\mathbf{c}$
and $\mathbf{\hat{c}}$, respectively. We define $d_H$ as the Hamming
distance between $\mathbf{c}$ and $\mathbf{\hat{c}}$. It is assumed
that the $d_H$ coded bits are interleaved such that they are placed
in distinct symbols. In addition, we know that the bit metrics
corresponding to the same coded bits between the pairwise errors are
the same. Based on the assumption and the knowledge,
(\ref{eq:PEP_original}) is re-written as
\begin{align}
\textrm{Pr} \left(\mathbf{c} \rightarrow \mathbf{\hat{c}} |
\mathbf{H}\right) = \textrm{Pr} \left( \sum_{k, d_H}
\min_{\mathbf{x} \in \xi_{c_{k'}}^{l,i}} \| \mathbf{r}_k -
\boldsymbol{\Gamma} \boldsymbol{\Theta} \mathbf{x} \|^2 \geq \right. \ifCLASSOPTIONtwocolumn \nonumber \\
\fi \left. \sum_{k, d_H} \min_{\mathbf{x} \in
\xi_{\hat{c}_{k'}}^{l,i}} \| \mathbf{r}_k - \boldsymbol{\Gamma}
\boldsymbol{\Theta} \mathbf{x} \|^2 \right)
\label{eq:PEP_for_different_codedbits}
\end{align}
where $\sum_{k, d_H}$ stands for the summation of the $d_H$ values
that correspond to the different coded bits between the codewords.

Let us define $\mathbf{\tilde{x}}_k$ and $\mathbf{\hat{x}}_k$ as
\begin{equation}
\begin{split}
\mathbf{\tilde{x}}_k = \arg \min_{\mathbf{x} \in \xi_{c_{k'}}^{l,i}}
\| \mathbf{r}_k - \boldsymbol{\Gamma} \boldsymbol{\Theta} \mathbf{x} \|^2 \\
\mathbf{\hat{x}}_k = \arg \min_{\mathbf{x} \in
\xi_{\bar{c}_{k'}}^{l,i}} \| \mathbf{r}_k - \boldsymbol{\Gamma}
\boldsymbol{\Theta} \mathbf{x} \|^2
\end{split}
\label{eq:arg_min}
\end{equation}
where $\bar{c}_{k'}$ is the complement of $c_{k'}$ in binary codes.
It is easily found that $\mathbf{\tilde{x}}_k$ is different from
$\mathbf{\hat{x}}_k$ since the sets that the $l^{th}$ symbols belong
to are disjoint, as can be seen from the definition of
$\xi_{c_{k'}}^{l,i}$. In the same manner, we see that $\mathbf{x}_k$
is different from $\mathbf{\hat{x}}_k$. With $\mathbf{\tilde{x}}_k$
and $\mathbf{\hat{x}}_k$, we get the following expression from
(\ref{eq:PEP_for_different_codedbits}) as
\begin{align}
\textrm{Pr} \left( \mathbf{c} \rightarrow \mathbf{\hat{c}} |
\mathbf{H} \right) = \textrm{Pr} \left( \sum_{k, d_H} \|
\mathbf{r}_k - \boldsymbol{\Gamma} \boldsymbol{\Theta}
\mathbf{\tilde{x}}_k
\|^2 \geq \right. \ifCLASSOPTIONtwocolumn \nonumber \\
\fi \left. \sum_{k, d_H} \| \mathbf{r}_k - \boldsymbol{\Gamma}
\boldsymbol{\Theta} \mathbf{\hat{x}}_k \|^2 \right).
\label{eq:alt_expression_PEP_diffbits}
\end{align}
Based on the fact that $\| \mathbf{r}_k - \boldsymbol{\Gamma}
\boldsymbol{\Theta} \mathbf{x}_k \|^2 \geq  \| \mathbf{r}_k -
\boldsymbol{\Gamma} \boldsymbol{\Theta} \mathbf{\tilde{x}}_k \|^2$
and the relation in (\ref{eq:detected_symbol}), equation
(\ref{eq:alt_expression_PEP_diffbits}) is upper-bounded by
\begin{align}
\textrm{Pr} (\mathbf{c} \rightarrow \mathbf{\hat{c}} | \mathbf{H})
\leq \textrm{Pr} \left( \beta \geq \sum_{k, d_H} \|
\boldsymbol{\Gamma} \boldsymbol{\Theta} (\mathbf{x}_k -
\mathbf{\hat{x}}_k) \|^2 \right) \label{eq:PEP_upperbounded}
\end{align}
where
\begin{equation*}
\beta = -\sum_{k, d_H} (\mathbf{x}_k - \mathbf{\hat{x}}_k)^H
\boldsymbol{\Theta}^H \boldsymbol{\Gamma} \mathbf{n}_k +
\mathbf{n}_k^H \boldsymbol{\Gamma} \boldsymbol{\Theta} (\mathbf{x}_k
- \mathbf{\hat{x}}_k).
\end{equation*}
Since $\beta$ is a zero mean Gaussian random variable with variance
$2N_0 \sum_{k, d_H} \|\boldsymbol{\Gamma} \boldsymbol{\Theta}
(\mathbf{x}_k - \mathbf{\hat{x}}_k)\|^2$,
(\ref{eq:PEP_upperbounded}) is replaced by the $Q$ function as
\begin{align}
\textrm{Pr} (\mathbf{c} \rightarrow \mathbf{\hat{c}} | \mathbf{H})
\leq Q\left( \sqrt \frac{\sum\limits_{k, d_H} \| \boldsymbol{\Gamma}
\boldsymbol{\Theta} (\mathbf{x}_k - \mathbf{\hat{x}}_k) \|^2}{2N_0}
\right). \label{eq:PEP_Q}
\end{align}
The numerator in (\ref{eq:PEP_Q}) is rewritten as
\begin{align}
\sum\limits_{k, d_H} \| \boldsymbol{\Gamma} \boldsymbol{\Theta}
(\mathbf{x}_k - \mathbf{\hat{x}}_k) \|^2 = \sum\limits_{k, d_H}
\sum\limits_{s=1}^S \lambda_s^2|d_{k,s}|^2 \ifCLASSOPTIONtwocolumn
\nonumber \\ \fi = \sum\limits_{s=1}^S \lambda_s^2 \sum\limits_{k,
d_H} |d_{k,s}|^2 \label{eq:Numer_FPMB}
\end{align}
where $\mathbf{d}_k = [d_{k,1} \cdots d_{k,S}]^T =
\boldsymbol{\Theta} (\mathbf{x}_k - \mathbf{\hat{x}}_k)$. Using an
upper bound to the $Q$ function, we calculate the average PEP as
\begin{align}
\textrm{Pr} (\mathbf{c} \rightarrow \mathbf{\hat{c}}) \leq E \left[
\exp \left( - \frac{\sum\limits_{s=1}^S \lambda_s^2 \sum\limits_{k,
d_H} |d_{k,s}|^2}{4 N_0} \right) \right] \label{eq:Avg_PEP_FPMB}.
\end{align}
In \cite{ParkICC09}, we have shown that equations with such form as
(\ref{eq:Avg_PEP_FPMB}) have a closed form expression of an upper
bound. We provide a formal description below.
\begin{theorem}
Consider the largest \mbox{$S \leq \min(N, M)$} eigenvalues $\mu_s$
of the uncorrelated central \mbox{$M \times N$} Wishart matrix that
are sorted in decreasing order, and a weight vector
\mbox{$\boldsymbol{\phi} = [\phi_1 \, \cdots \, \phi_S]^T$} with
non-negative real elements. In the high signal-to-noise ratio
regime, an upper bound for the expression $E [ \exp (-\gamma
\sum_{s=1}^S \phi_s \mu_s ) ]$ which is used in the diversity
analysis of a number of MIMO systems is
\begin{align*}
E\left[ \exp \left( - \gamma \sum\limits_{s=1}^S \phi_s \mu_s
\right) \right] \leq \zeta \left( \phi_{min} \gamma
\right)^{-(N-\delta+1)(M-\delta+1)}
\end{align*}
where $\gamma$ is signal-to-noise ratio, $\zeta$ is a constant,
$\phi_{min} = \min \{ \phi_1, \, \cdots, \, \phi_S \}$, and $\delta$
is the index to the first non-zero element in the weight vector.
\label{theorem:E_PEP}
\end{theorem}
\begin{IEEEproof}
See \cite{ParkICC09}.
\end{IEEEproof}
By calculating the weight vector whose $s^{th}$ element is $\sum_{k,
d_H} |d_{k,s}|^2$, we evaluate the diversity order of a given
system. In particular, if the constellation precoder is designed
such that
\begin{align}
|d_{k,1}|^2 = | \boldsymbol{\theta}_1^T ( \mathbf{x}_k -
\mathbf{\hat{x}_k} ) |^2
> 0, \forall (\mathbf{x}_k, \mathbf{\hat{x}}_k
)\label{eq:condition_full_diversity}
\end{align}
where $\boldsymbol{\theta}_1^T$ is the first row vector of the
unitary precoding matrix $\boldsymbol{\Theta}$, we see that
$\sum_{k, d_H} |d_{k,1}|^2
> 0$, resulting in the full diversity order of $NM$. Therefore,
(\ref{eq:condition_full_diversity}) is a sufficient condition for
the full diversity order of BICMB-FP.

\subsection{BICMB with Partial Precoding} \label{sec:PPMB}

The bit metrics in (\ref{eq:ML_bit_metrics_PPMB}) lead to the PEP
calculation as
\begin{align}
\textrm{Pr} \left(\mathbf{c} \rightarrow \mathbf{\hat{c}} |
\mathbf{H}\right) = \textrm{Pr} \left( \tau_1 \geq \tau_2 \right)
\label{eq:PEP_for_different_codedbits_PPMB_case3}
\end{align}
where
\begin{align*}
\tau_1 &= \sum_{k, d_{H,p}} \min_{\mathbf{x} \in
\psi_{c_{k'}}^{l,i}} \| \mathbf{r}_{k}^p - \boldsymbol{\Gamma}_p
\boldsymbol{\tilde{\Theta}} \mathbf{x} \|^2 + \sum_{k,d_{H,n}}
\min_{x \in \chi_{c_{k'}}^{l,i}} | r_{k,l} - \lambda_{\hat{l}} x |^2
\end{align*}
\begin{align*}
\tau_2 &= \sum_{k, d_{H,p}} \min_{\mathbf{x} \in
\psi_{\bar{c}_{k'}}^{l,i}} \| \mathbf{r}_{k}^p -
\boldsymbol{\Gamma}_p \boldsymbol{\tilde{\Theta}} \mathbf{x} \|^2 +
\sum_{k, d_{H,n}} \min_{x \in \chi_{\bar{c}_{k'}}^{l,i}} | r_{k, l}
- \lambda_{\hat{l}} x |^2
\end{align*}
and $\sum_{k, d_{H,p}}$, $\sum_{k, d_{H,n}}$ stand for the summation
over the $d_{H,p}$ and $d_{H,n}$ bit metrics corresponding to the
different coded bits carried on the subchannels in $\mathbf{b}_p$
and $\mathbf{b}_n$, respectively. By using the appropriate system
input-output relations, the PEP is written as
\begin{align}
\textrm{Pr} \left(\mathbf{c} \rightarrow \mathbf{\hat{c}} |
\mathbf{H} \right) = \textrm{Pr} \left( \beta \geq \kappa \right)
\label{eq:PEP_upperbounded_PPMB3}
\end{align}
where $\beta = \beta_p + \beta_n$,
\begin{align*}
\beta_p = -\sum_{k, d_{H,p}} (\mathbf{x}_{k, \mathbf{b}_p} -
\mathbf{\hat{x}}_{k, \mathbf{b}_p})^H \boldsymbol{\tilde{\Theta}}^H
\boldsymbol{\Gamma}_p \mathbf{n}_{k}^p + \ifCLASSOPTIONtwocolumn
\nonumber \\ \fi \left(\mathbf{n}_{k}^p\right)^H
\boldsymbol{\Gamma}_p \boldsymbol{\tilde{\Theta}} (\mathbf{x}_{k,
\mathbf{b}_p} - \mathbf{\hat{x}}_{k, \mathbf{b}_p}),
\end{align*}
\begin{align*}
\beta_n = - \sum_{k, d_{H,n}} \lambda_{\hat{l}} (x_{k, l} -
\hat{x}_{k,l})^*n_{k,l} + \lambda_{\hat{l}} (x_{k, l} -
\hat{x}_{k,l}) n_{k,l}^*,
\end{align*}
and
\begin{align*}
\kappa = \sum_{k, d_{H,p}} \| \boldsymbol{\Gamma}_p
\boldsymbol{\tilde{\Theta}} \left( \mathbf{x}_{k, \mathbf{b}_p} -
\mathbf{\hat{x}}_{k, \mathbf{b}_p} \right) \|^2 +
\ifCLASSOPTIONtwocolumn \nonumber \\ \fi \sum_{k, d_{H,n}} |
\lambda_{\hat{l}} \left( x_{k,l} - \hat{x}_{k,l} \right) |^2.
\end{align*}
Since $\beta$ in (\ref{eq:PEP_upperbounded_PPMB3}) is a Gaussian
random variable with zero mean and variance $2N_0 \kappa$, the PEP
can be expressed in a similar way as (\ref{eq:PEP_Q}) with the
$Q$-function. In addition, if we define $\sigma$ as
\begin{equation}
\sigma = \sum_{r=1}^P \lambda_{b_p(r)}^2 \sum_{k, d_{H,p}}
|\hat{d}_{k, r}|^2 + d^2_{min}\sum_{r=1}^{S-P} \lambda_{b_n(r)}^2
\alpha_{b_n(r)} \label{eq:kappa_lowerbound}
\end{equation}
where $\mathbf{\hat{d}}_k = [\hat{d}_{k,1} \, \cdots \,
\hat{d}_{k,P}]^T = \boldsymbol{\tilde{\Theta}} \left( \mathbf{x}_{k,
\mathbf{b}_p} - \mathbf{\hat{x}}_{k, \mathbf{b}_p} \right)$, and
$\alpha_s$ is the number of times the $s^{th}$ subchannel is used
corresponding to $d_{H,n}$ bits under considertion, then we can see
that $\sigma \geq \kappa$. Finally, the average PEP is calculated as
\begin{align}
\textrm{Pr} \left(\mathbf{c} \rightarrow \mathbf{\hat{c}} \right)
\leq E \left[ \frac{1}{2} \exp \left( - \frac{\sigma}{4 N_0} \right)
\right]. \label{eq:Avg_PEP_PPMB_case3}
\end{align}

To determine the diversity order from $\sigma$, we need to find the
index to indicate the first non-zero element in an ordered composite
vector which consists of $\sum_{k, d_{H,p}} | \hat{d}_{k,r}|^2$ and
$\alpha_{b_n(r)}$ as in Theorem \ref{theorem:E_PEP}. If $d_{H,p} =
0$, the first summation part of $\sigma$ vanishes. In this case, the
first index is
\begin{equation}
\delta = \min\{  s : \alpha_{s}
> 0 \textrm{ for } s \in \{b_n(1), \, \cdots, \, b_n(S-P)\}\}.
\label{eq:delta_PPMB}
\end{equation}
In the other case of $d_{H,p} > 0$, we see that $\mathbf{x}_{k,
\mathbf{b}_p}$ and $\mathbf{\hat{x}}_{k, \mathbf{b}_p}$ are
obviously different for the same reason as in the previous section.
If the constellation precoder satisfies the sufficient condition of
(\ref{eq:condition_full_diversity}), the term with
$\lambda_{b_p(1)}^2$ always exists in $\sigma$. Therefore, $\delta$
for the case of $d_{H,p} > 0$ is $\delta = \min(b_p(1), \delta')$
where $\delta'$ is obtained in the same way as
(\ref{eq:delta_PPMB}).

\textit{Example of Determining Diversity Order : }

In this example, we employ $4$-state $1/2$-rate convolutional code
with generator polynomials $(5, 7)$ in octal representation in $N =
M = S = 3$ system. Two types of spatial interleavers are used to
demonstrate the different results of the diversity order. A
generalized transfer function of BICMB with the specific spatial
interleaver and convolutional code provides the $\alpha$-vectors for
all of the pairwise errors, whose element indicates the number of
times the stream is used for the erroneous bits \cite{ParkICC09}. In
particular, due to the fact that $d_{H,p} = \sum_{r=1}^P
\alpha_{b_p(r)}$ and $d_{H,n} = \sum_{r=1}^{S-P} \alpha_{b_n(r)}$
where $\alpha_s$ is the $s^{th}$ element of the $\alpha$-vector, the
generalized transfer function is also useful in the analysis of
BICMB-PP. Hence, we rewrite the transfer functions of the systems
from \cite{ParkICC09}, where $a$, $b$, and $c$ are the symbolic
representation of the $1^{st}, 2^{nd}, 3^{rd}$ stream. The spatial
interleaver used in $\mathbf{T}_1$ is a simple rotating switch on
$3$ streams. For $\mathbf{T}_2$, the $u^{th}$ coded bit is
interleaved into the stream $s_{\mathrm{mod}(u-1, 18)+1}$ where
$s_1$ = $\cdots$ = $s_6$ = $1$, $s_7$ = $\cdots$ = $s_{12}$ = $2$,
$s_{13}$ = $\cdots$ = $s_{18}$ = $3$ and $\mathrm{mod}$ is the
modulo operation. Each term represents the $\alpha$-vector, and the
powers of $a$, $b$, $c$ indicate the elements of $\alpha$-vector.
\begin{align}
\label{eq:Transfunc_S3R1_2} \mathbf{T}_1 &= Z^5(a^2 b^2 c + a^2 b
c^2 + a b^2 c^2) + Z^6(a^3 b^2 c + \cdots) \nonumber\\
&+ Z^7(2 a^3 b^3 c + 2 a^3 b^2 c^2 + 2 a^2 b^3 c^2 + \cdots) \\
&+ Z^8(a^5 b^3 + b^5 c^3 + a^3 c^5 + \cdots) + \cdots \nonumber
\end{align}
\begin{align}
\label{eq:Transfunc_S3R1_2_diffdemux}
\mathbf{T}_2 &=Z^5 (a^5 + a^3 b^2 + a^2 b^3 +\nonumber\\
&\quad\qquad b^5 + a^3 c^2 + b^3 c^2 + a^2 c^3 + b^2 c^3 + c^5)\nonumber\\
&+ Z^6(a^4 b^2 + 3 a^3 b^3 + a^2 b^4 + a^4 c^2 + 3 a^2 b^2 c^2 + \\
&\quad\qquad b^4 c^2 + 3 a^3 c^3 + 3 b^3 c^3 + a^2 c^4 + b^2 c^4) +
\cdots \nonumber
\end{align}

Consider the case $\mathbf{b}_p = [1 \, 2]$. We see that all of
$\alpha$-vectors of $\mathbf{T}_1$ show $d_{H,p}
> 0$, leading to $\delta = 1$. Therefore, the
diversity order of the $\mathbf{T}_1$ BICMB-PP system with
$\mathbf{b}_p = [1 \, 2]$ achieves the full diversity order while
BICMB without constellation precoding \cite{ParkICC09}, or PPMB
without bit-interleaved coded modulation (BICM) loses the full
diversity order \cite{park-2009_arxiv} \cite{ParkGlobecom09}.
However, $\mathbf{T}_2$ has $[0 \, 0 \, 5]$ which shows $d_{H,p} =
0$, resulting in $\delta = 3$. Therefore, the diversity order of the
$\mathbf{T}_2$ BICMB-PP system with $\mathbf{b}_p = [1 \, 2]$ does
not achieve the full diversity order.

The same analysis for $\mathbf{b}_p = [1 \, 3]$ results in the
diversity order of $9$, and $[2 \, 3]$ results in $4$ for the
transfer function $\mathbf{T}_1$. Similarly, both of $[1 \, 3]$ and
$[2 \, 3]$ result in the diversity of $4$ for $\mathbf{T}_2$. As a
consequence, we find that proper selection of the subchannels for
precoding, as well as the appropriate pattern of the spatial
interleaver, is important to achieve the full diversity order of
BICMB-PP.

%% file: results.tex
\section{Simulation Results} \label{sec:results}

Monte-Carlo simulations were performed to verify the diversity
analysis in Section \ref{sec:div_analysis}. Throughout the
simulations, we used the precoding matrices in
\cite{park-2009_arxiv}, \cite{ParkGlobecom09} which meet the
sufficient condition to achieve the full diversity order of
(\ref{eq:condition_full_diversity}). Fig.
\ref{fig:2x2_BICMBvsBICFPMB} depicts the simulation result for $2
\times 2$, $3 \times 3$, and $4 \times 4$ BICMB and BICMB-FP with
$64$-state convolutional code punctured from $1/2$-rate mother code
with generator polynomials $(133, 171)$ in octal representation. In
\cite{ParkICC09}, we showed the maximum achievable diversity order
of BICMB with an $R_c$-rate convolutional code is $(N-\lceil S \cdot
R_c \rceil+1)(M-\lceil S \cdot R_c \rceil+1)$. In this example, the
maximum achievable diversity order of the three BICMB systems is
$1$. However, Fig. \ref{fig:2x2_BICMBvsBICFPMB} shows that BICMB-FP
achieves the full diversity order for any code rate. Fig.
\ref{fig:3x3_4Q_BICPPMB} depicts the simulation results of BICMB-PP
given in the example of Section \ref{sec:PPMB}. The diversity orders
of the BICMB systems, $\mathbf{T}_1$ and $\mathbf{T}_2$ are $4$ and
$1$, respectively. We see that the simulation results match the
analysis in \ref{sec:PPMB}.

\ifCLASSOPTIONonecolumn
\begin{figure}[!m]
\centering \includegraphics[width =
0.6\linewidth]{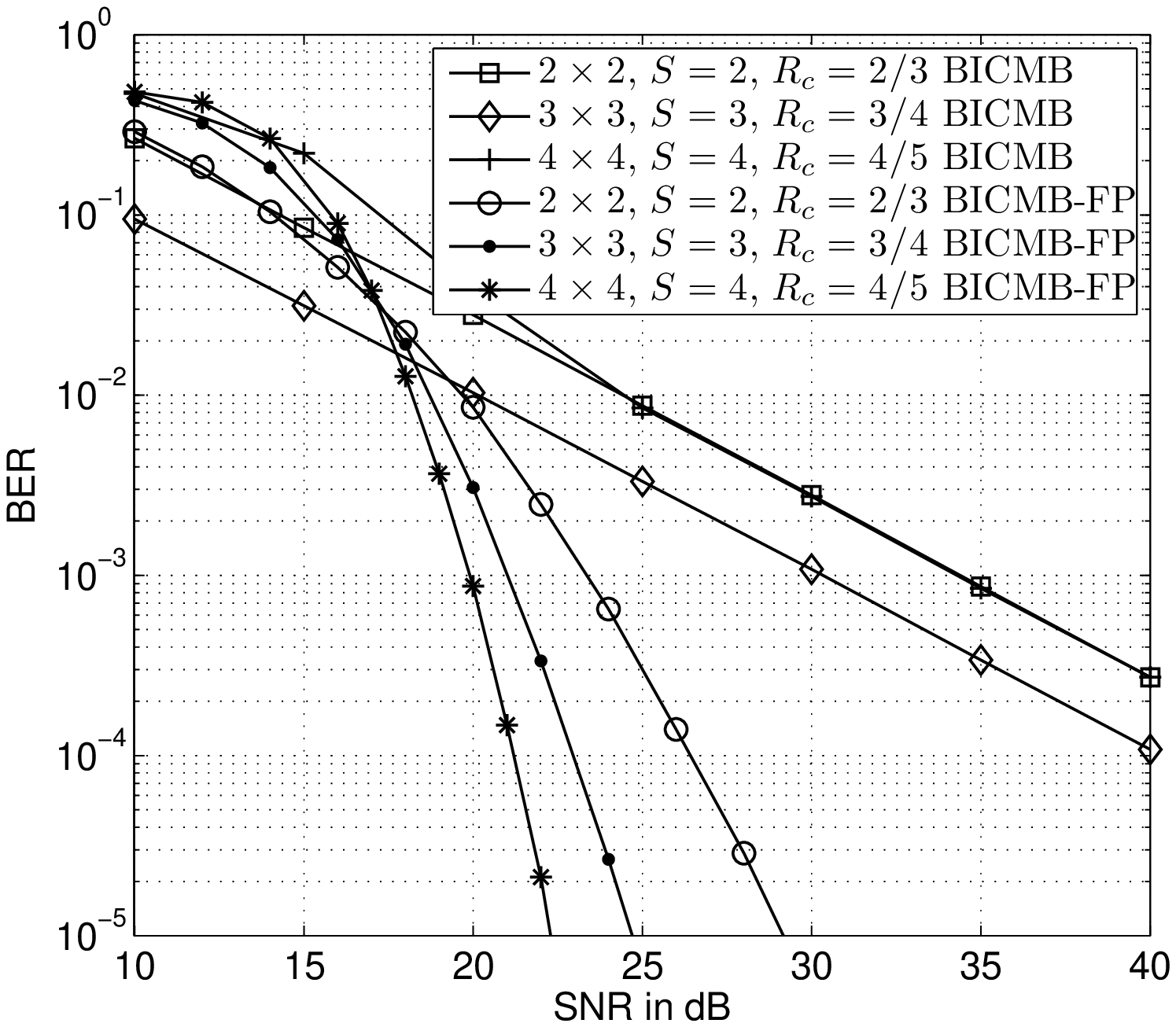} \caption{BER comparison between
BICMB and BICMB-FP with $16$-QAM, and $64$-state punctured
convolutional code.} \label{fig:2x2_BICMBvsBICFPMB}
\end{figure}
\else
\begin{figure}[!t]
\centering \includegraphics[width =
0.8\linewidth]{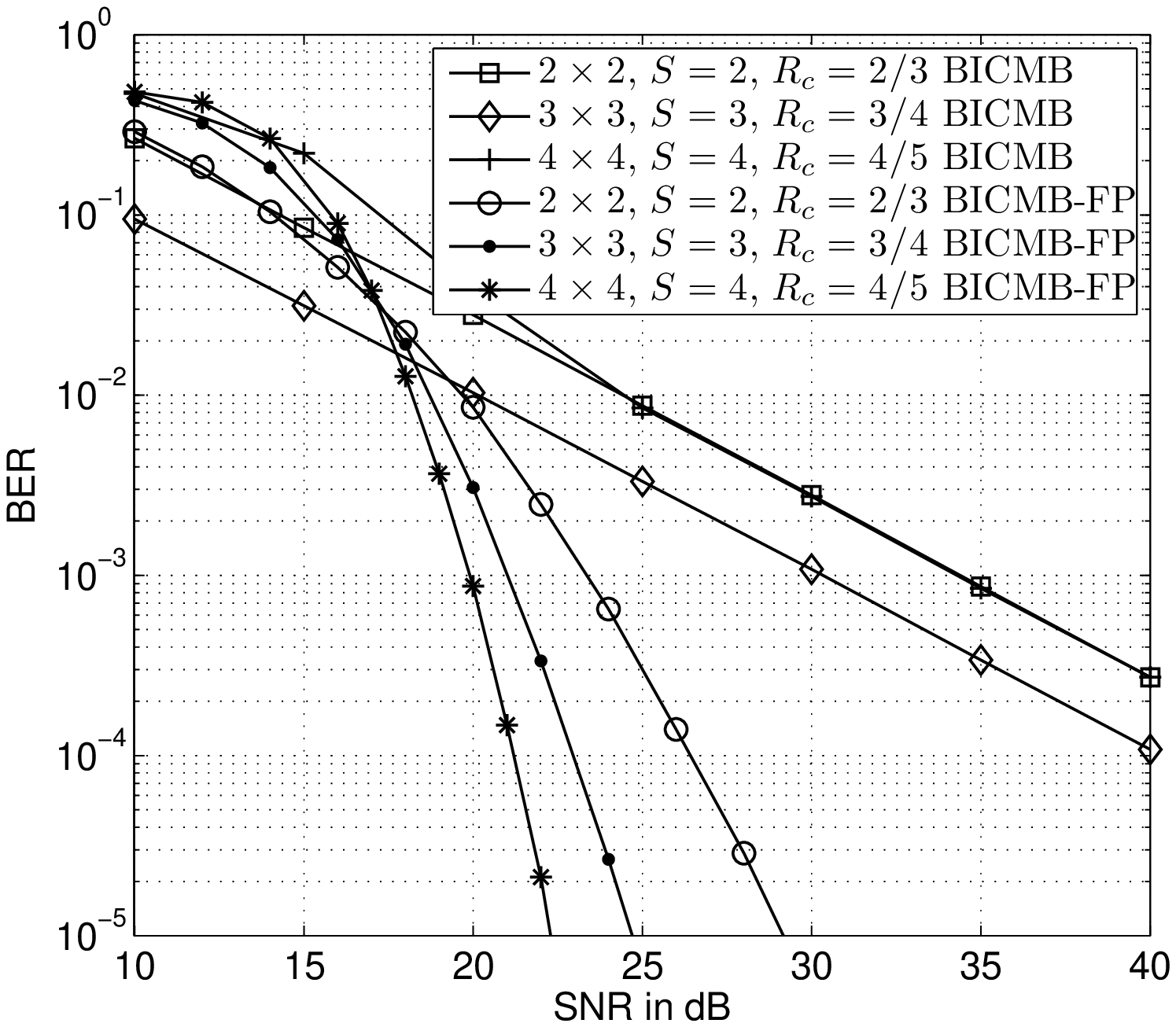} \caption{BER comparison between
BICMB and BICMB-FP with $16$-QAM, and $64$-state punctured
convolutional code.} \label{fig:2x2_BICMBvsBICFPMB}
\end{figure}
\fi

\ifCLASSOPTIONonecolumn
\begin{figure}[!m]
\centering \includegraphics[width =
0.6\linewidth]{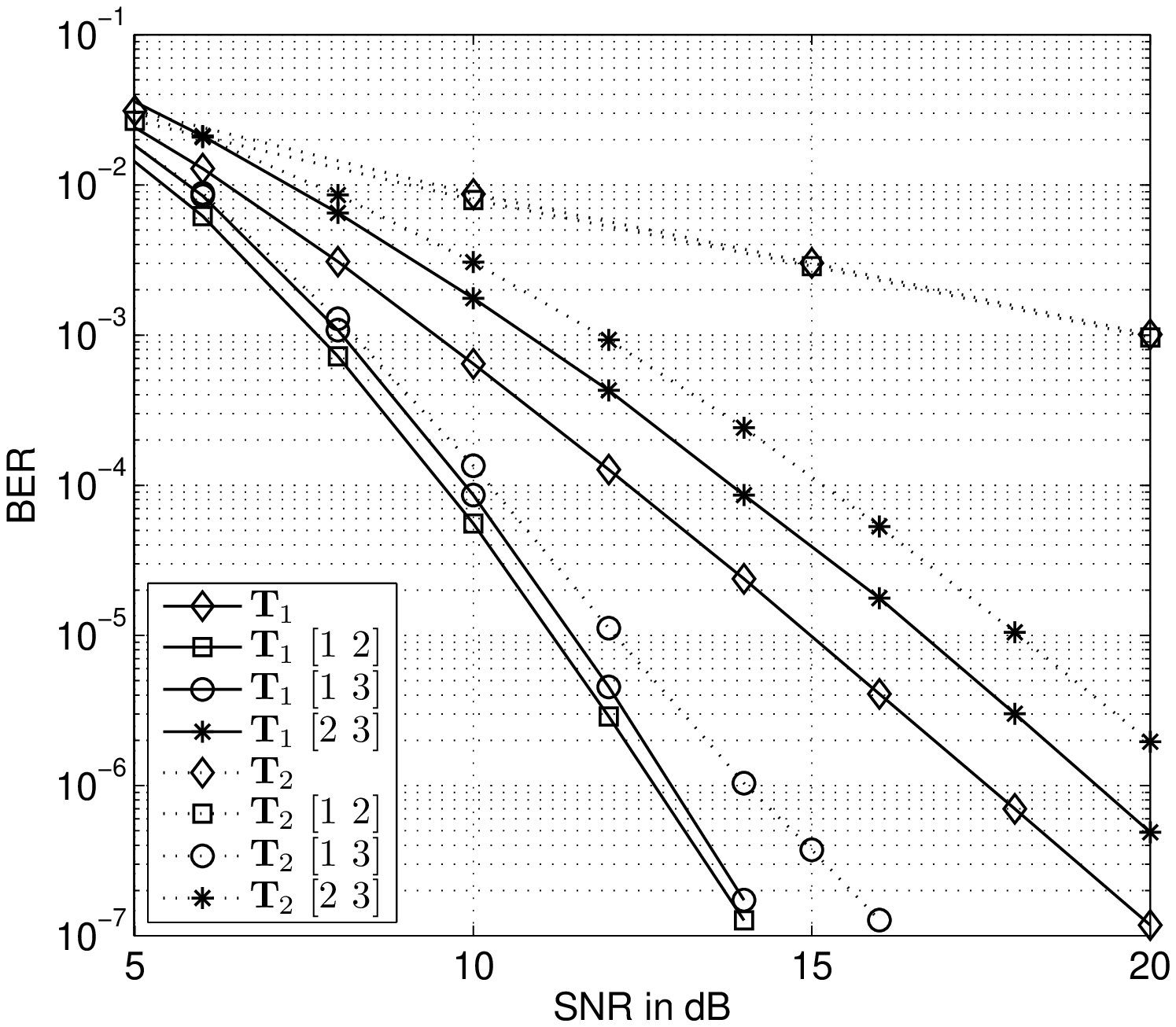} \caption{BER vs. SNR for BICMB-PP
with $3 \times 3$ $S=3$, $4$-QAM, and $4$-state $1/2$-rate
convolutional code.} \label{fig:3x3_4Q_BICPPMB}
\end{figure}
\else
\begin{figure}[!t]
\centering \includegraphics[width =
0.8\linewidth]{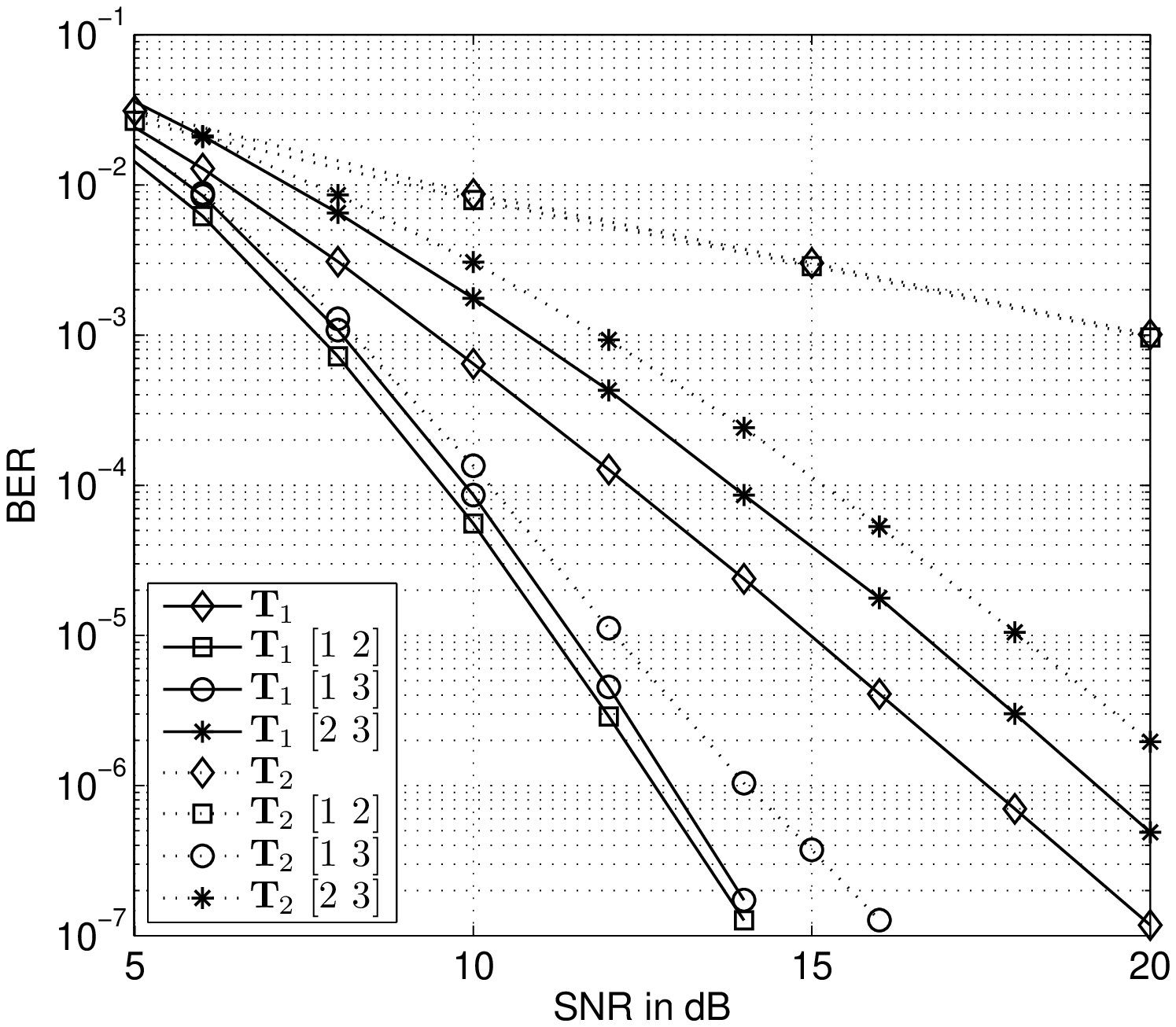} \caption{BER vs. SNR for for
BICMB-PP with $3 \times 3$ $S=3$, $4$-QAM, and $4$-state $1/2$-rate
convolutional code.} \label{fig:3x3_4Q_BICPPMB}
\end{figure}
\fi

To compare coding gain between BICMB and BICMB-PP that achieve the
full diversity order, we show in Fig. \ref{fig:4x4_BICMB_PP} the BER
performances with the $4 \times 4$ $S=4$ systems. The used generator
polynomials of $64$-state, $1/2$ and $1/4$-rate convolutional codes
are $(133, 171)$ and $(117, 127, 155, 171)$ in octal representation,
respectively \cite{FrengerCOML99}. As shown in the figure, BICMB
combined with constellation precoding shows larger coding gain with
a large number of antennas and at the higher transmission rate.

\ifCLASSOPTIONonecolumn
\begin{figure}[!m]
\centering \includegraphics[width = 0.6\linewidth]{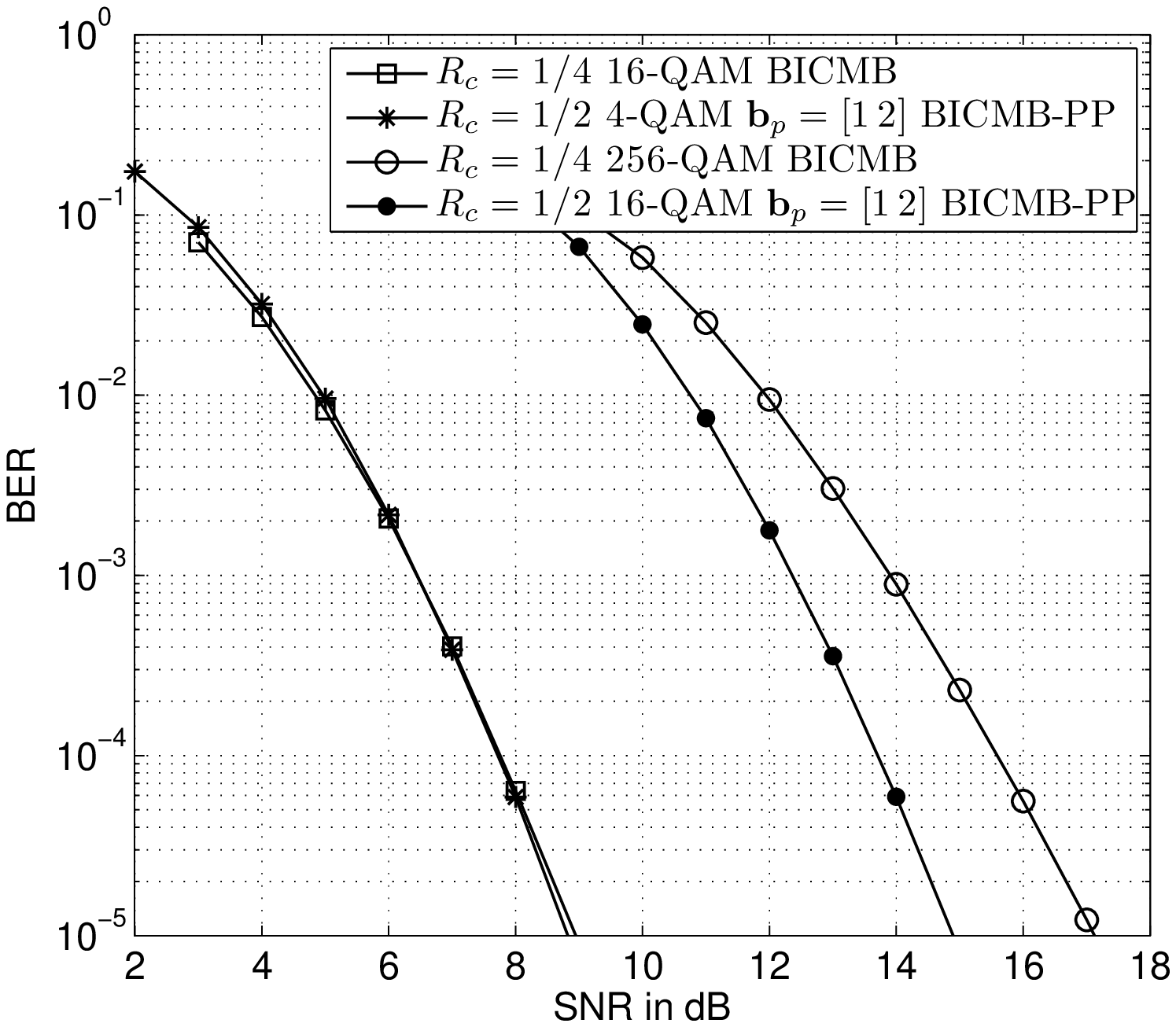}
\caption{BER comparison between $4 \times 4$ $S=4$ BICMB and
BICMB-PP at $4$ and $8$ bits/channel use.} \label{fig:4x4_BICMB_PP}
\end{figure}
\else
\begin{figure}[!t]
\centering \includegraphics[width = 0.8\linewidth]{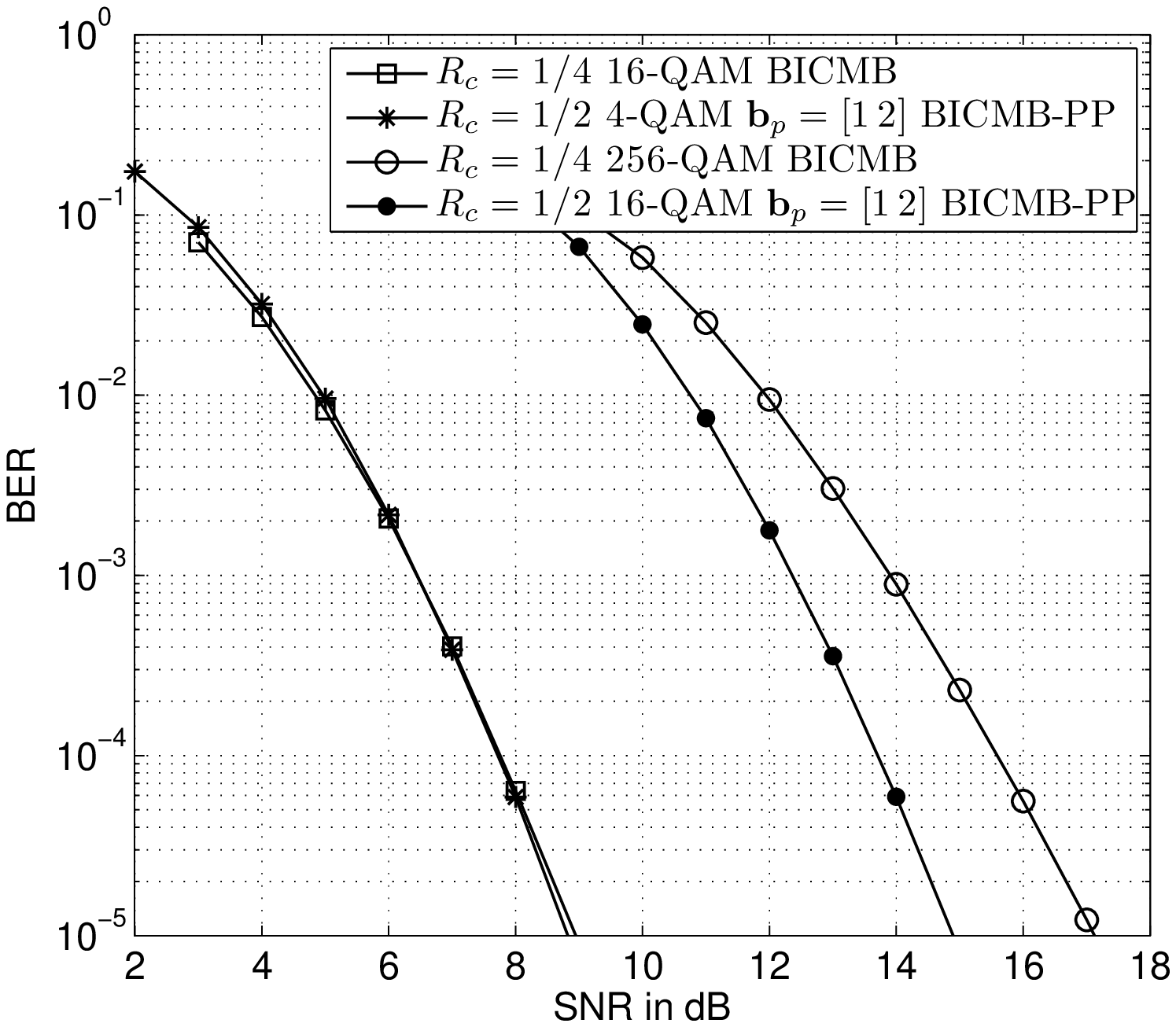}
\caption{BER comparison between $4 \times 4$ $S=4$ BICMB and
BICMB-PP at $4$ and $8$ bits/channel use.} \label{fig:4x4_BICMB_PP}
\end{figure}
\fi

%% file: conclusion.tex
\section{Conclusion} \label{sec:conclusion}

We investigated the diversity order of BICMB combined with the
constellation precoding scheme, by calculating pairwise error
probability. Using the analysis, we presented the resulting
diversity order of the given examples. The analysis can be used to
determine the precoding configuration from the given BICMB
implementation to get the full diversity order. We provided
simulation results that proves the analysis. In addition, the
simulation showed that BICMB-PP outperforms BICMB with a large
number of antennas and at the higher transmission rate.